\def\epem{$e^+e^-$ }
\def\as{$\alpha_s$ }
\def\jpsi{J/\psi}
\def\be{\begin{eqnarray}}
\def\ee{\end{eqnarray}}
\def\app#1#2#3{{\it Act. Phys. Pol. }{\bf B #1} (#2) #3}
\def\apa#1#2#3{{\it Act. Phys. Austr.}{\bf#1} (#2) #3}

\def\npb#1#2#3{{\it Nucl. Phys. }{\bf B #1} (#2) #3}
\def\plb#1#2#3{{\it Phys. Lett. }{\bf B #1} (#2) #3}
\def\prd#1#2#3{{\it Phys. Rev. }{\bf D #1} (#2) #3}
\def\prl#1#2#3{{\it Phys. Rev. Lett. }{\bf #1} (#2) #3}

\def\jet#1#2#3{{\it JETP Lett. }{\bf #1} (#2) #3}
\def\zpc#1#2#3{{\it Z. Phys. }{\bf C #1} (#2) #3}

\documentstyle[epsfig,subfigure,12pt,here]{article}
\begin{document}
%
  
\renewcommand{\thefootnote}{\fnsymbol{footnote}}
\begin{flushright}
  {\bf TTP96-53\footnote{The complete paper, including
      figures, is also 
      available via anonymous ftp at
      ftp://ttpux2.physik.uni-karlsruhe.de/, or via www at\\
      http://www-ttp.physik.uni-karlsruhe.de/cgi-bin/preprints/}}\\
{\bf hep-ph/9612372}\\
November 1996
\end{flushright}
\vspace*{15mm}
\begin{center}
{\bf \large Heavy Flavor Production and Decay}\\[1cm]
{\bf\large J.~H.~K\"uhn\footnote{Supported by BMBF contract
    057KA92P.\\
Invited talk presented at the ``Cracow International Symposium on
    Radiative Corrections'', Cracow, Poland, 1996.}}\\[5mm]
{\em Institut f\"ur Theoretische Teilchenphysik, 
         Universit\"at Karlsruhe,\\ D-76128 Karlsruhe, Germany\\[2mm]}
\end{center}
\thispagestyle{empty}
\begin{abstract}
Recent theoretical results on
heavy flavor production and decay in the framework of perturbative QCD
are reviewed. This includes 
calculations for top production at hadron colliders, inclusive
charmonium production and the comparison between the singlet and octet
mechanisms.  Predictions for heavy flavor production in \epem
annihilation will be discussed in some detail, covering both the
threshold and the high energy region.  The first results in 
NLO for heavy flavor decays will also be
reviewed.
\end{abstract}
\newpage
%


\section{Introduction}
Heavy flavor production and decay have developed into benchmark
reactions for perturbative QCD.  The large energy scale inherent in
most of these reactions allows for a separation between hard and soft
momentum transfers.  The former can be treated perturbatively, the
nonperturbative matrix elements which encode the remaining information
can either be determined experimentally, or integrated out by
considering sufficiently inclusive information such that perturbation
theory alone is adequate. 

Significant progress has been achieved recently in a number of
topics.  The predictions for top production at hadron colliders
have been scrutinized by
several authors.  In particular the role of soft gluon resummation has
been emphasised and the \as dependence explored (Section 2).
Inclusive charmonium production at hadron and \epem colliders has been
studied theoretically and experimentally.  A fairly complex picture
seems to emerge, with different mechanisms playing a role in various
reactions (Section 3).  
The inclusive cross section for heavy flavor production in \epem
annihilation has been studied in a variety of papers.  Far above
threshold an expansion in $m^2/s$ is adequate and has been
successfully applied to $Z$ decays to bottom quarks, or to charm
production just below the $b\bar b$ threshold.  For a prediction
above, but relatively close to threshold a different strategy has been
employed, which is based on a combination of analytical and numerical
methods.  For an adequate treatment of top quark production in the
threshold region its large decay rate and the interplay between gluon
radiation from the production and the decay process must be taken into
account.  These topics will be reviewed in section 4.  The leading QCD
correction to weak decays of heavy flavors have been evaluated quite
some time ago.  Results are available for the rate, the spectrum and
for angular distributions.  To match the level of precision claimed by
the proponents of the Heavy Quark Effective Theory, next to leading
order predictions are required from perturbation theory.  First steps
into this direction have been made and will be reviewed in section 5.

\section{Top production in hadronic collisions}

\begin{figure}[htbp]
\centerline{\epsfig{figure=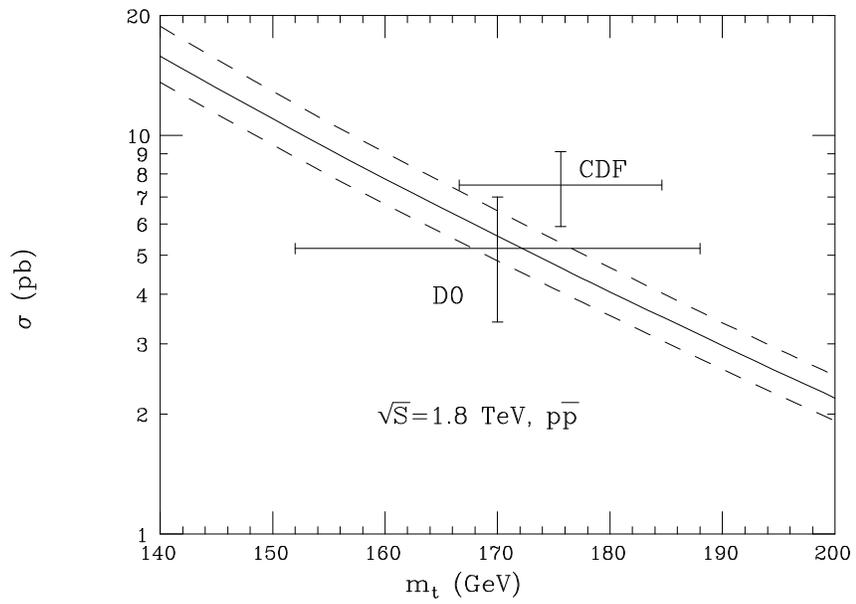,width=0.9\textwidth,clip=}}
\caption{\label{fig:1} Top cross section at the Tevatron at
  $\protect\sqrt{S}=2\;$TeV (from \protect\cite{Catani1}).}
\end{figure}

The theoretical framework and the (semi-) analytical results for the
top production cross section in NLO have been developed nearly a
decade ago \cite{EQ,nason1}. The predictions for $\sqrt{s}=1.8$ GeV
and $m_t$ = 180 GeV from various authors are listed in
Table~\ref{table:history}. 
\begin{table}[htb]
\begin{center}
\begin{tabular}{|l|l|}
\hline
  &  $\protect\sigma$ [pb] \\ \hline
Altarelli et al. \protect\cite{nason1}  & $3.52$ (DFLM) 
\\
                                          & $4.10$ (ELHQ)
\\ \hline
Laenen et al. \protect\cite{Laenen} & 
           $ \left.\begin{array}{ll} 3.5 &\quad(\mu^2=4m^2)\\
                                   3.8  &\quad(\mu^2=m^2)\\
                                   4.05 &\quad(\mu^2=m^2/4)
                  \end{array}
\right\}$ MRSD
\\ \hline
\hline
Resummation &   \\ \hline
Laenen et al. \cite{Laenen}   &  $
            \left.\begin{array}{c} 3.86\\4.21\\4.78\end{array}
\right\} $ vary $\mu_0$
\\ \hline
Berends et al. \cite{Berends} & 4.8 central value
\\ \hline
Berger et al. \cite{Berger} & 4.8 ``principal value res.''
\\ \hline
Catani et al. \cite{Catani} & $4.05^{+0.62}_{-0.52}$
\\ \hline
\end{tabular}
\end{center}
\caption{History of predictions for the production cross section for
$\protect\sqrt{s} =1.8$ TeV and $m_t = 180$ GeV.}
\label{table:history}
\end{table}
For $m_t$ = 175 GeV the cross section increases by about 0.7 pb.  The
uncertainty in the factorization and renormalisation scale leads to an
uncertainty of roughly 10\%.  Recently the issue of soft gluon
resummation has been raised. The original arguments \cite{Laenen,Berger}
leading to a large positive shift of roughly 10\% have been refuted in
\cite{Catani}.  No consensus has yet been reached on the magnitude of
these effects. Increasing \as from the nominal value of around 0.11,
which has been frequently used in these calculations, to
0.120 leads to an increase by about 5\%.  Within the combined
uncertainties theory and experiment are in very good agreement
(Fig.~\ref{fig:1}). 

All these calculations are based on a perturbative treatment of the
threshold region.  In principle one should, however, incorporate the
leading terms of order $\pi \alpha_s/\beta$.  The resulting
modifications are small for $t\bar t$ in a color octet which is the
dominant configuration at the TEVATRON (see section 2.2.2 in
\cite{top}).  

\section{Inclusive Charmonium Production}

High energy hadron-hadron and $e^-p$ colliders are charmonium
factories.  A variety of production mechanisms have been discussed in
the literature.  Contributing with different relative strengths in the
various reactions they can be disentangled only through a systematic
study of different processes.  In particular the question of color
singlet versus octet production has stimulated a number of detailed
investigations.

\begin{figure}[h]
\begin{center}
  \leavevmode
\centerline{\epsfig{figure=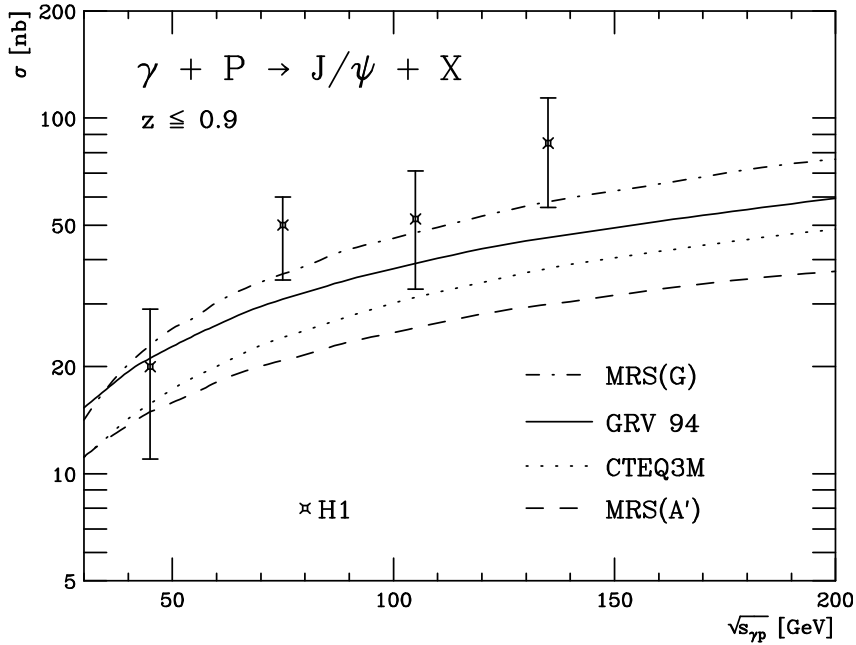,width=0.9\textwidth,clip=}}
  \vspace*{-8mm}
\end{center}
\caption{Comparison between theoretical prediction for the energy
  dependence of the inelastic cross section ($z\le 0.9$) for $\jpsi$
  photoproduction (J. Steegborn, private communication, based on
  \protect\cite{Kraemer}) and recent data from the H1 Collaboration.}
\label{fig:2}
\end{figure}

Inelastic $\jpsi$ production in photon-photon reactions provides a
relatively clean testing ground.  The dominant subprocess at the
parton level 
\be
\gamma+g\to \jpsi + g
\ee
can produce directly a ($c\bar c$) color singlet state.
Incorporating also the one loop perturbative corrections
\cite{Kraemer}, 
satisfactory
agreement between theory and experiment is observed for the $\jpsi$
energy distributions and the total production cross section as well
(Fig.~\ref{fig:2}). 

This success of the color singlet model (CSM) (where quarkonium
(color singlet!) states are required to be produced through a purely
perturbative mechanism) is in marked contrast
with its failure in purely hadronic collisions.  The dominant
subprocesses in the CSM are based on the conversion of a virtual gluon
into $\jpsi$ or $\chi_J$ plus two or one gluon respectively.  The
combination of additional powers of \as with the small phase space
gives rise to sizable suppression factors.  This perturbative
treatment of soft gluon radiation may be inadequate and an alternative
approach has been advocated in \cite{cit1}.  The cross section for
charmonium production is decomposed into a sum of terms consisting of
the cross section for ($c\bar c$) states in a specific angular
momentum and color state times the nonperturbative matrix element of
an operator characterizing the conversion probability into $\jpsi$:
\be
\sigma(p\bar p\to \jpsi + x) = \sum_n 
\sigma(p\bar p\to c\bar c(n))\times \langle {\cal O}^{\jpsi}_n\rangle.
\ee
These matrix elements are effectively free parameters to be determined
in different experiments.  This approach is closely related in its
spirit to the color evaporation model formulated a long time ago; it
provides, however, a more firm theoretical formulation.  Adjusting the
parameters appropriately, a satisfactory description of the data is
obtained.

The clean initial state configuration typical for \epem annihilation
is ideal to investigate the relative importance of different
production mechanisms.  Two distinctly different situations have been
considered: high energy reactions like $Z$ decays with large event
rates available at LEP  and alternatively the 10 GeV region that can be
explored at present at CESR or in the near future at the $B$-meson
factories.  Three mechanisms have been identified at which contribute
in the high energy region with comparable rates. 
The reaction \cite{keu}
\be
Z\to \jpsi c\bar c+X 
\ee
requires the production of two $c\bar c$ pairs with a rate
proportional  to $\alpha_s^2|R(0)|^2$.  The second mechanism
\cite{keu2} 
is the
splitting of a virtual gluon in a color octet $c\bar c$:
\be
Z\to q\bar q (c\bar c)_8
\ee
with the subsequent nonperturbative conversion of $(c\bar c)_8$ into
$\jpsi$.  The rate for this mechanism is proportional to
$\alpha_s^2\langle{\cal O}^8\rangle$ where the second factor
characterizes the nonperturbative matrix element.  The third,
color singlet, contribution \cite{hag}
\be
Z\to q\bar q \jpsi gg
\ee
is strongly suppressed by the factor $\alpha_s^4|R(0)|^2$ and,
furthermore, by the small phase space.  The branching ratios of the
three reactions are given by $0.8\cdot 10^{-4}$, $1.9\cdot 10^{-4}$,
$0.2\cdot 10^{-4}$,  respectively.  The total inclusive rate is
reasonably consistent with the observations by the OPAL collaboration
\cite{OPAL}
of $(1.9\pm 0.7 \pm 0.5 \pm 0.5)\cdot 10^{-4}$. However, a statement
about the dominance of any of these processes seems premature.  The
analysis of $\jpsi$ energy and momentum distributions, however, could
help to settle this issue.

\begin{figure}[h]
\begin{center}
  \leavevmode
\centerline{\epsfig{figure=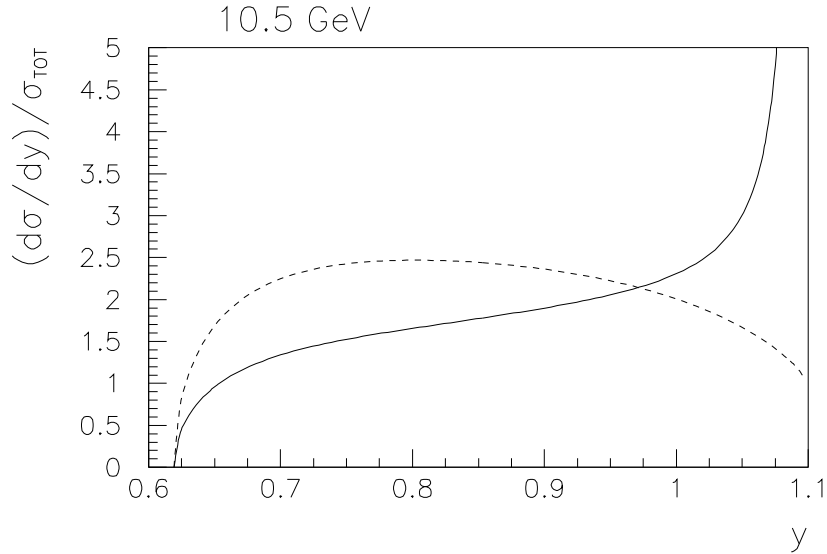,width=0.9\textwidth,clip=}}
  \vspace*{-5mm}
\end{center}
\caption{Energy distribution for inclusive $\jpsi$ production in
  $e^+e^-$ annihilation at 10.5 GeV. Solid curve: $\alpha_s(M^2_{gg})$,
  dashed curve $\alpha_s(M^2_{\psi})$.}
\label{fig:3}
\end{figure}

\begin{figure}[h]
\begin{center}
  \leavevmode
\centerline{\epsfig{figure=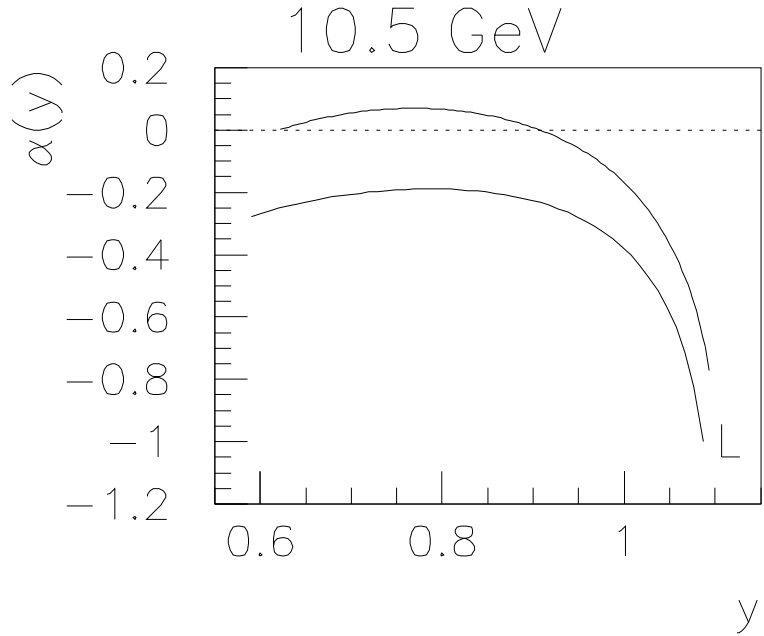,width=0.9\textwidth,clip=}}
  \vspace*{-10mm}
\end{center}
\caption{Coefficient $\alpha(y)$ characterizing the angular
  distribution of $\jpsi$'s (L: longitudinally polarized $\jpsi$'s
  only). }
\label{fig:4}
\end{figure}

Also $B$ meson factories and CESR give rise to a large sample of
events with prompt $\jpsi$ production.  Two mechanisms have been
proposed which might well describe complementary kinematical regions.
The leading process in the CSM
\be 
e^+ e^- \to \jpsi +gg
\ee
is proportional to $\alpha_s^2|R(0)|^2$.  It leads to a three body
final state and hence to a continuous energy distribution
(Fig.~\ref{fig:3}). 
Predictions for the rate, the angular and the momentum distribution
and the polarization can be found in \cite{Driesen}.  The alternative
approach \cite{bra} is based on ``color octet production'', $e^+e^-\to (c\bar
c)_8+g$. The rate is of order \as and multiplied by a 
nonperturbative matrix
element.  The $\jpsi$ energy is essentially fixed at
$E_{max}=(s+m^2_\psi)/(2\sqrt{s})$.  The angular distribution is
proportional to $(1+\cos^2\theta)$.  These features are identical to
the predictions of the ``color evaporation model'' \cite{Fritsch}
formulated a long time ago.  An excess of $\jpsi$ at this special
kinematical point with the predicted angular distribution would be a
strong indication for this ``octet mechanism''. The angular
distribution of the $\jpsi$ in the CSM is of the form $1+\alpha(y)
\cos^2\theta$ where $\alpha(y)$ depends on $y\equiv E_\psi /E_{Beam}$
and  approaches roughly $-0.8$ at the endpoint
(Fig.~\ref{fig:4}).
This difference will be crucial in disentangling the two mechanisms.

\section{Heavy flavor production in \epem annihilation}

\subsection{$Z\to b\bar b$}
Experimental studies of various partial and of the total $Z$ decay
rate  have been performed recently with a new level of sophistication.
The relative error in $\Gamma_b$ has been lowered to about $0.5 \cdot
10^{-2}$ corresponding to $\delta \Gamma_b \approx 2.5 $ MeV, the
uncertainty in the total decay rate which is also influenced by
$\Gamma_b$ amounts to about 3 MeV.  In comparison with $\Gamma_d$ or
$\Gamma_u$  two important differences have to be taken into account
for $\Gamma_b$.  The first, relatively straightforward aspect is
related to the bottom mass.  In Born approximation the correction from
the phase space suppression of the axial part of the rate is predicted
to be $-6m_b^2/M_Z^2$ corresponding to $-4$ MeV. In \cite{CK} it has
been demonstrated that this number is drastically modified by QCD
corrections.  The bulk of these, the large logarithms, can be absorbed
by reexpressing the result in terms of the running mass thus reducing
the correction to $-1.6$ MeV.  (For a detailed discussion and further
references see \cite{Report}.)  The second contribution to the $Z\to
b\bar b$ decay has its origin in the double triangle diagrams with two
gluon intermediate states.
It is present for the axial rate only.  The contribution of order
$\alpha_s^2$ was calculated quite some time ago for arbitrary
$m_t^2/M_Z^2$.  Formally it is proportional to $\ln m_t^2/M_Z^2$ and
thus seems to diverge in the limit of large $\ln
m_t^2/M_Z^2$. However, additional logarithms of $m_t^2/\mu^2$ are
induced by the running of $\alpha_s$ which have to be controlled at
the same time.  The structure of leading logs was analysed in
\cite{Chet}, the constant terms of $\alpha_s^3$ in
\cite{ChetTar}. The combined effect of order $\alpha_s^2$ and
$\alpha_s^3$ from these ``singlet terms'' amounts $\delta \Gamma_b =
-1.8$ MeV.  It is clear that the sum of mass and singlet terms must be
taken into consideration in any precision analysis.  

\subsection{Intermediate energies}
\begin{figure}[h]
\begin{center}
\vspace*{-3cm}
\mbox{\epsfig{file=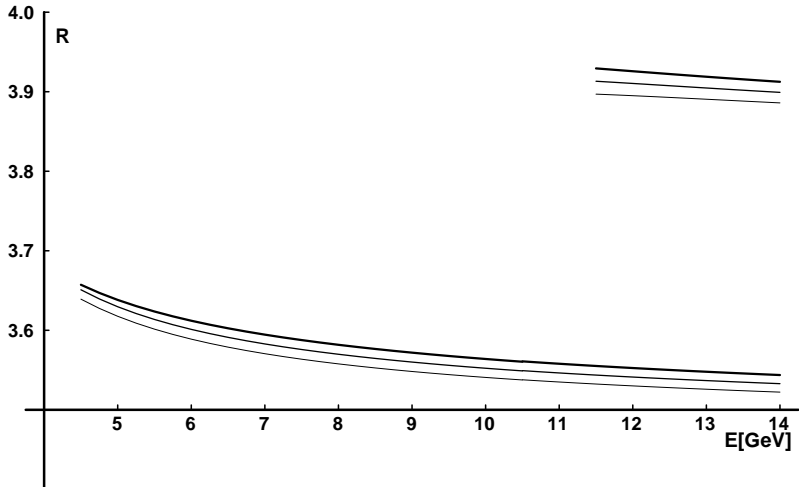,width=10.5cm}}
\end{center}
\vspace*{-5cm}
\caption{\label{fig:6}
The ratio $R(s)$  below and above
the ${\rm b}$ quark production threshold at $10.5$ GeV
for  $\alpha_s(M_{{Z}}) = 0.120, \, 0.125$ and $0.130$.
The contributions from  light quarks ($u$, $d$, $s$, $c$) 
and the bottom quark are displayed separately.}
\end{figure}

The $Z$ decay rate is well described in the massless approximation
plus terms of order $m_b^2/M_Z^2$.  However, for a prediction at lower
energies, an increasing series of terms in the $m^2/s$ expansion is
needed. The comparison between the complete calculation and a limited
number of terms in the $m^2/s$ expansion indicates that the first three
terms are sufficient to describe the cross section from high energies
down to $s\approx 8m^2$. With this motivation in mind the quartic
terms of order $\alpha_s^2$ have been calculated in \cite{Quart}.
In this way  an adequate prediction between roughly 14 GeV and $M_Z$
is available for $b\bar b$ production, and similarly for $c\bar c$
production from roughly 5 to 6
GeV up to the bottom quark threshold \cite{Rhad}
(Fig.~\ref{fig:6}).

In view of the large statistics available at CESR and at a future
$B$-meson factory a detailed theoretical study has been performed in
\cite{Teubner} which demonstrates the potential for this potentially most
precise and clean determination of $\alpha_s$.  

\subsection{The NLO calculation for arbitrary $m^2$ and s}
A few GeV above charm, bottom, or top threshold measurements can in
principle be performed at a $\tau$-charm factory, at a $B-$meson
factory and a future linear collider.  With a relative momentum of the
quarks exceeding for instance 3 GeV perturbative QCD should be
applicable also in this region.  It is, therefore, desirable to push
the theoretical prediction as close as possible towards the threshold.
The two-loop calculation has been performed more than 40 years ago
\cite{Kallen}. The imaginary part of those three-loop diagrams which
originate from massless fermion loop insertions in the gluon
propagator (``double bubble diagrams'') were calculated analytically
in \cite{HKT}.  Real and imaginary parts of the purely gluonic
correction (and of the double bubble diagrams) were calculated in a
semianalytical approach \cite{ChKS} that will be sketched in this
subsection.

The polarization function can be written in the form
\be
\lefteqn{\Pi = \Pi^{(0)} + {\alpha_s\over \pi}\Pi^{(1)}}
\nonumber \\
&&
+ \left({\alpha_s\over \pi}\right)^2
\left(
C_F^2 \Pi_A + C_FC_A \Pi_{NA} + C_FTn_l \Pi_{l} + C_FT \Pi_{F}
\right) 
\ee
where $n_l$ denotes the number of light quark species.
Each one of the $\Pi_j$ is analytical in the complex $q^2$ plane
with a cut from $4m^2$ to $+\infty$.  For small $q^2$ they
can be expanded in a Taylor series
\be
\Pi(q^2,m^2) = \sum_{n>0}C_n \left( {q^2\over 4m^2}\right)^n
\ee
The renormalization condition $\Pi(q^2=0,m^2)=0$ has already been
implemented. The evaluation of the Taylor coefficients amounts to the
calculation of three loop tadpole integrals with an increasing number
of mass insertions -- up to 16 for $C_8$ which is the present limit
for the evaluation with the help of algebraic programs. 

\begin{figure}
 \begin{center}
 \begin{tabular}{c}
   \epsfxsize=9cm
   \leavevmode
   \epsffile[110 330 460 520]{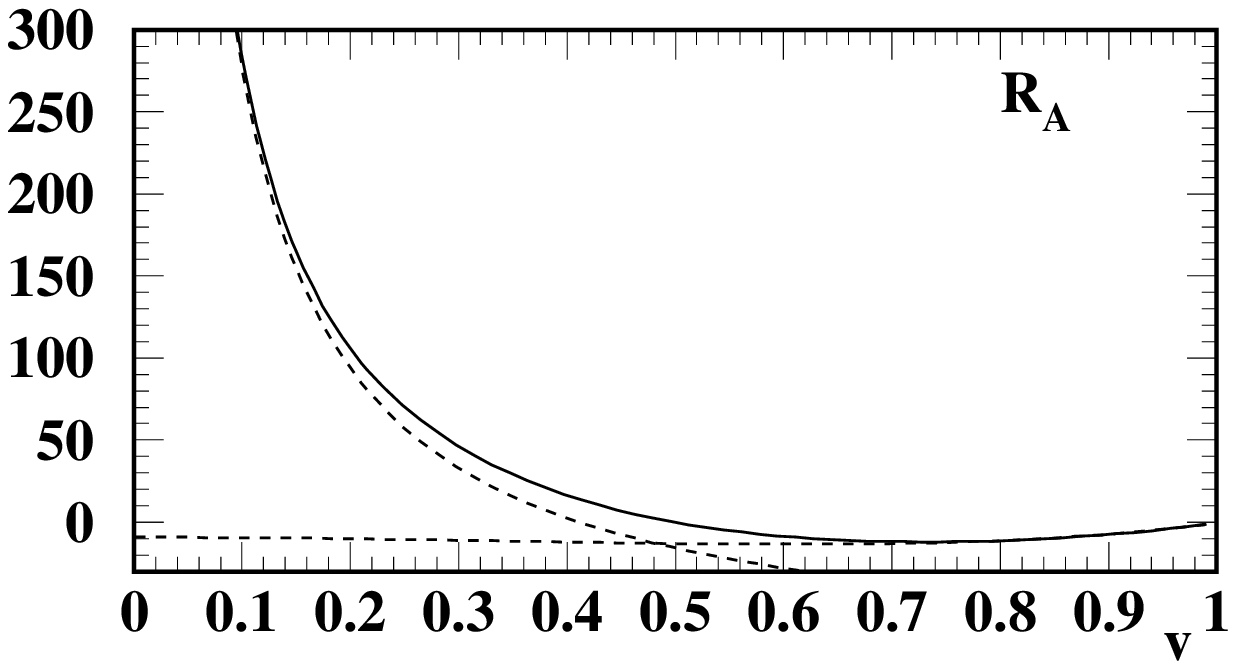}  
   \\
   \epsfxsize=9cm
   \leavevmode
   \epsffile[110 330 460 520]{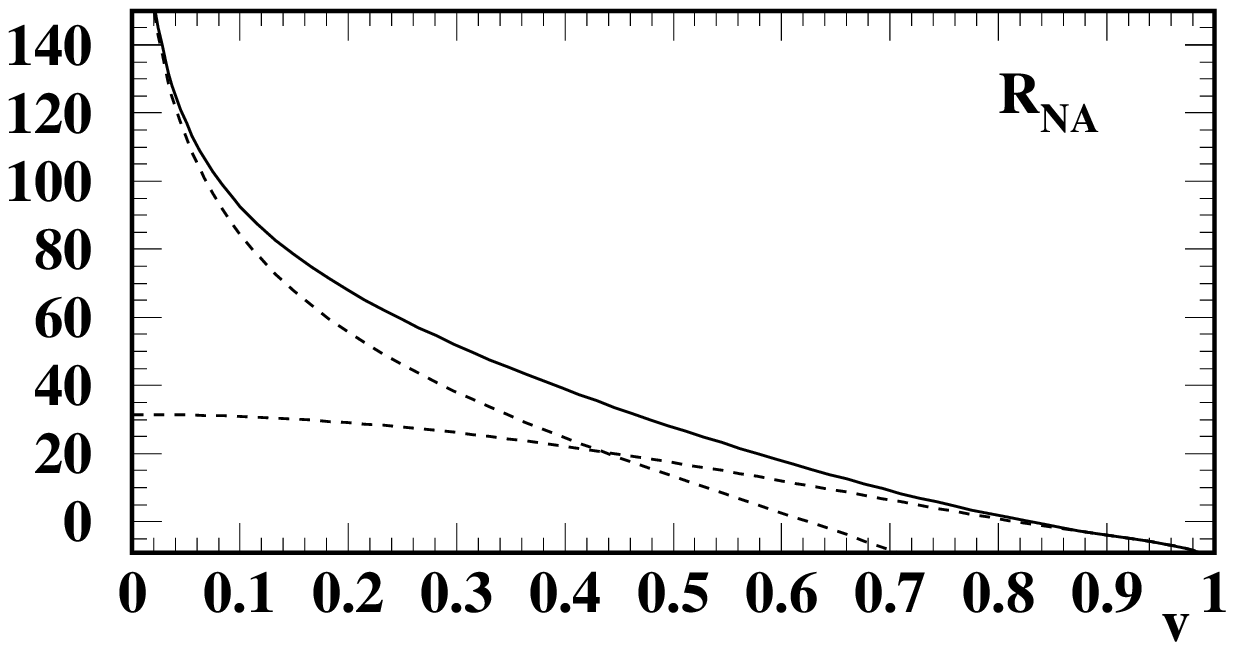} 
   \\
   \epsfxsize=9cm
   \leavevmode
   \epsffile[110 330 460 520]{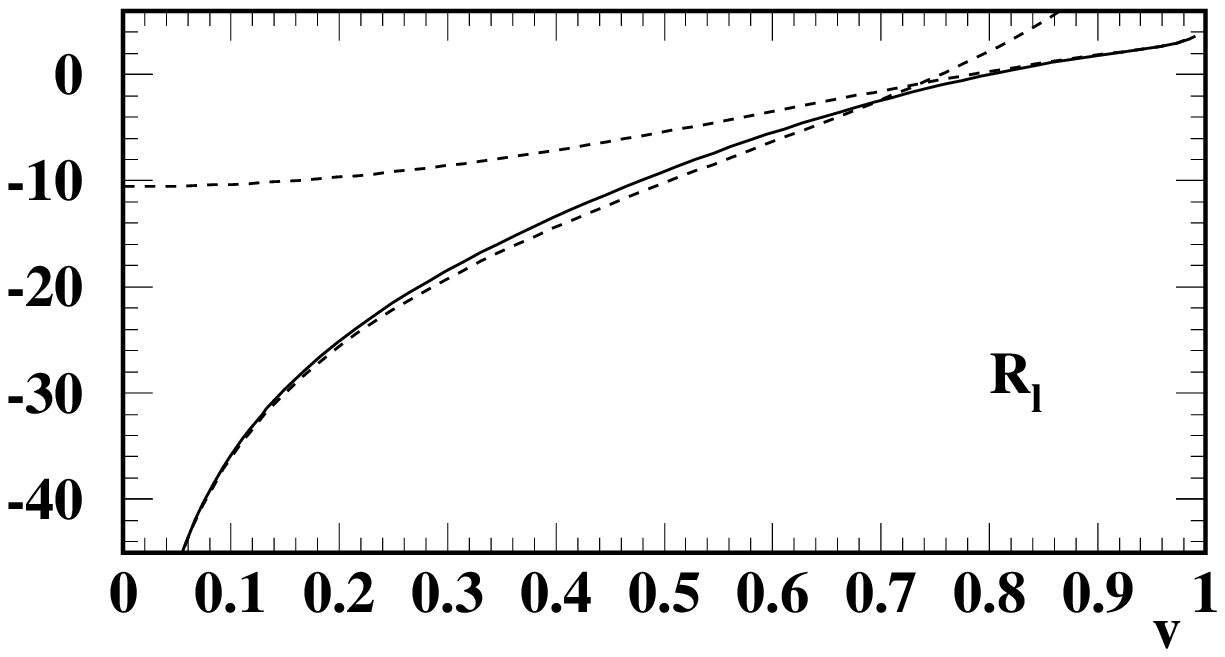}
 \end{tabular}
 \caption{\label{fig:71} Complete results
                          plotted against
                          $v=\protect\sqrt{1-4m^2/s}$. The
                          high energy approximation includes the $m^4/s^2$ 
                          term.}
 \end{center}
\end{figure}

In the large $q^2$ region a similar expansion can be performed. For
this case the expansion has been performed up to terms
of order $(m^2/q^2)^0$ and $(m^2/q^2)^1$.  Additional information can
be obtained about the behavior close to threshold.  Leading and
subleading terms can be deduced from the influence of the Coulomb
potential in the nonrelativistic region, combined with the knowledge
about the logarithmic corrections of the perturbative QCD potential.
To extend the range of convergence from $q^2<4m^2$ to the full
analyticity domain an appropriate variable transformation has to be
performed. The data from the three kinematical regions are finally
integrated in a Pad\'e approximation which leads to stable results for
$\Pi(q^2)$ and $R(s)$ at the same time.  The result for the three
dominant 
pieces are shown in Fig.~\ref{fig:71} where it is compared
to the leading terms close to the threshold and to the high energy
approximation. 
\subsection{Toponium and top quarks in the threshold region}
\begin{figure}[ht]
  \begin{center}
    \leavevmode
    \epsfxsize=12.cm
    \epsffile[110 265 465 560]{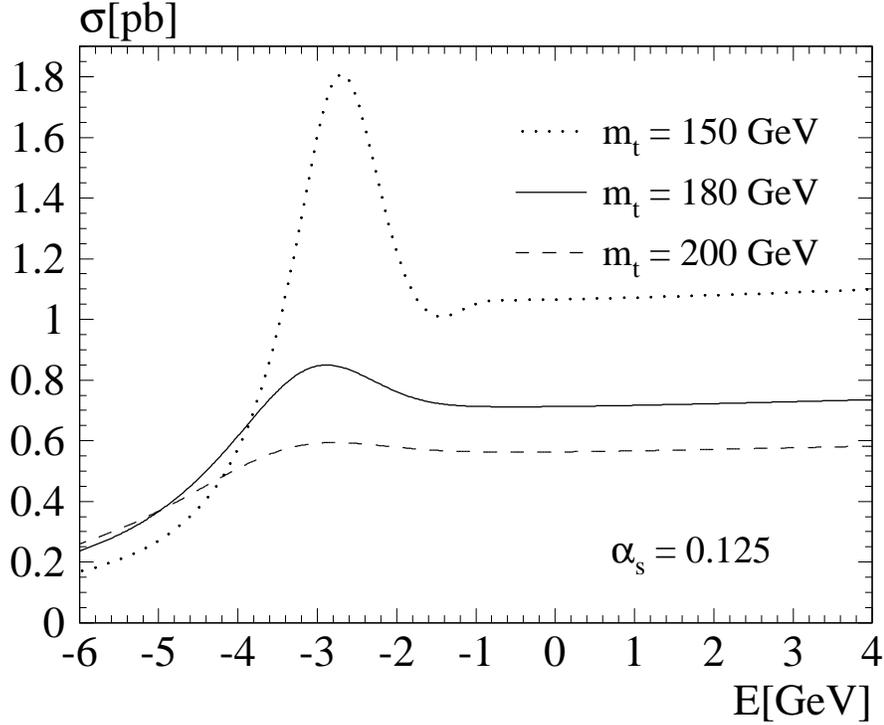}
    \hfill
\caption{Total cross section as function of $E=\protect\sqrt{s}-2m_t$
for three values of the top quark mass.}
\label{fig:8}
  \end{center}
\end{figure}

Top quarks were treated as stable particles in the previous section.
Although adequate away from the threshold, this approximation is
inadequate in the ``would-be'' toponium region.  For a mass of the
top quark around 175 GeV a decay rate $\Gamma_t\approx 1.5$ GeV is
predicted, corresponding to a toponium width of 3 GeV. The resonances
are thus completely dissolved \cite{rep26,rep43}, 
and the individual peaks are merged
into a step function like threshold cross section.
Quarkonium physics ceases to exist.  The large decay rate introduces,
however, a cutoff which eliminates all nonperturbative aspects of the
interquark potential. Large momentum tails beyond 
\be
P_{cut} \approx \sqrt{2m_t\Gamma_t} \approx 24 GeV
\ee
or, alternatively, distances above 
\be
r\approx 0.01 {\rm fm}
\ee
are irrelevant for the description of the $t\bar t$ system
\cite{FK,JezKT,JezT,Sum}. The impact of the large rate is clearly
visible in Fig.~\ref{fig:8}.
The predictions for three different top masses $m_t=150$ GeV, 180 GeV,
and 200 GeV corresponding to $\Gamma_t=$ 0.81 GeV, 1.57 GeV, and  2.24
GeV
demonstrate the strong influence of $\Gamma_t$ on the shape of
the cross section.  The shape is furthermore significantly modified
by initial state radiation and the spread in the beam energy.

\begin{figure}[ht]
  \begin{center}
    \leavevmode
    \epsfxsize=12.cm
    \epsffile[15 210 580 635]{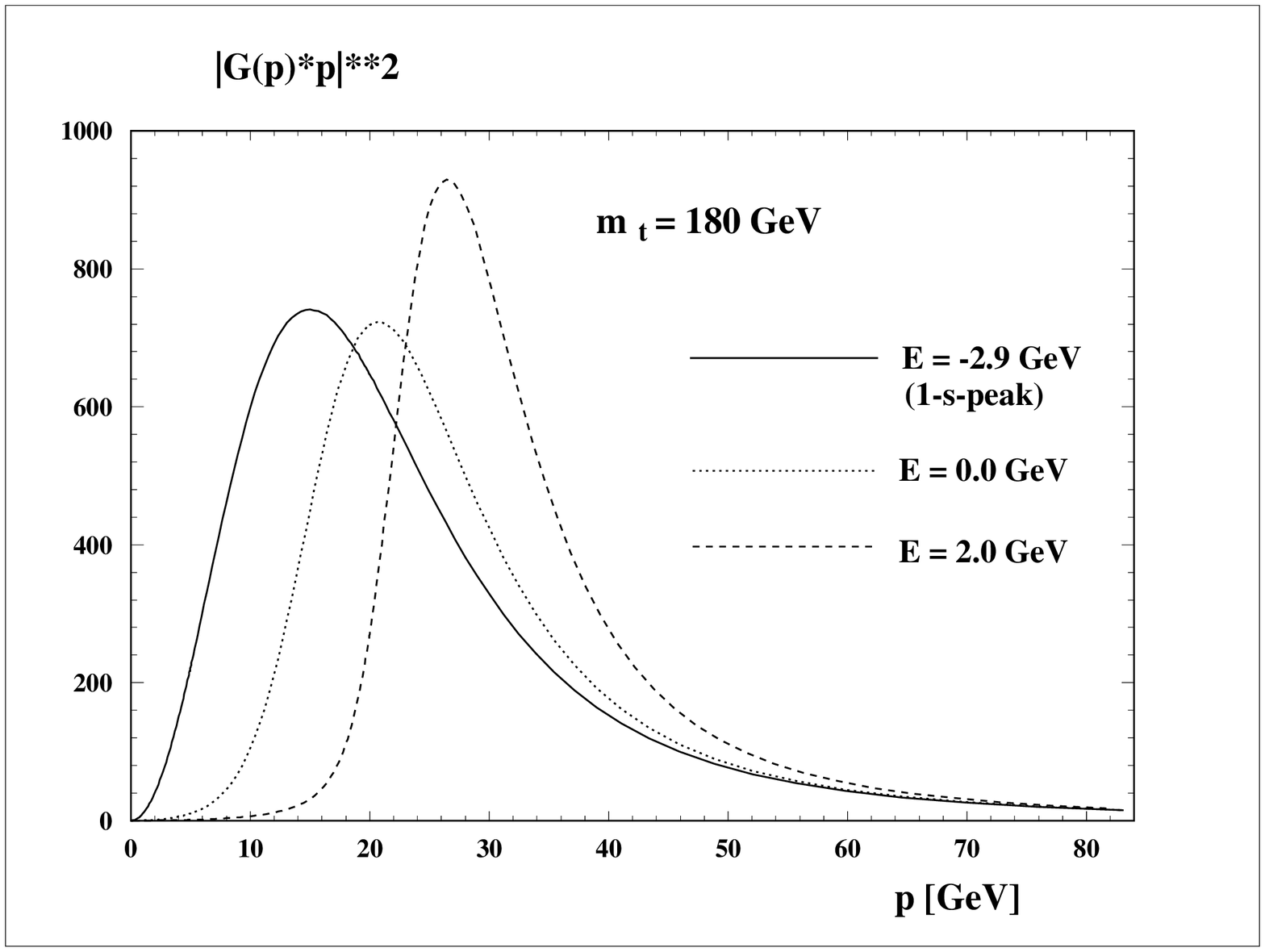}
    \hfill
\caption{Momentum distribution of top quarks for three different cms
energies.}
\label{fig:9}
  \end{center}
\end{figure}

Additional information is encoded in the momentum distribution of top
quarks, the ``Fermi motion'' which can be traced through the decay
products $W+b$. This distribution is essentially equivalent to the
square of the wave function in momentum space and can, for unstable
particles, be evaluated \cite{JezKT,JezT,Sum} 
with Green's function techniques
(Fig.~\ref{fig:9}).
Various experimental studies have demonstrated the potential of a
linear collider to determine $m_t$ to a precision of perhaps even 200
MeV by measuring the cross section and the momentum distribution
simultaneously.

 Highly polarized top quarks are required for a variety
of precision studies of top decays.  In the threshold region this is
easily achieved.  In fact, even with unpolarized beams top
quarks are longitudinally polarized (with a polarization around
$-0.4$)  as a consequence of the
nonvanishing axial part of the neutral current.
Longitudinally
polarized beams lead to a fully polarized sample of top quarks.

Another step in complication is achieved by considering the
interference between the dominant $S$ and the suppressed $P$ wave
contributions.
The relative size of these effects is of order
$\beta\sim 0.1$. It leads to a forward--backward asymmetry
\cite{top122} and
furthermore to an angular dependent quark polarization
perpendicular to the beam direction \cite{top123}. A detailed
discussion of these 
effects, in particular of the role of the normal polarization and of
rescattering corrections, can be found in \cite{us}.
The small polarization of top quarks normal to the production plane is
a particularly sensitive measure of the interquark potential.

Additional complications are introduced through the rescattering
\cite{us,Sumino}
between $b$ quark jets and the spectator, and by relativistic
corrections \cite{JezT} 
of order $\alpha_s^2$.  
These effects will be important for the quantitative comparison
between theory and experiment and the extraction of a precise value
for $m_t$, $\Gamma_t$ and $\alpha_s$ from threshold studies.

\section{Towards NLO in Heavy Flavor Decays}

Semileptonic weak decays of bottom mesons and top quarks are
particularly clean probes of the fundamental properties of quarks,
their masses and mixing angles.  Decay rates are, however, influenced
by QCD effects, a large part of which can be calculated in PQCD.
Leading order corrections to practically all quantities of interest
are available: for the decay rate of charmed and bottom quark from
\cite{CabMai} and for top quarks from \cite{jk}.  Lepton decay spectra
have been calculated in \cite{jkdis}, the energy distribution of
hadrons in \cite{had}.  Leptons from the decays of polarized quarks
exhibit a nontrivial angular distribution \cite{pol,cjkk94} and even lepton
mass effects have been incorporated in these calculations
\cite{tau,Motyka}. A compact summary of most of these QCD corrections
can be found in \cite{sum}.

Different techniques to determine the degree of $b$ or top
polarization have been investigated in \cite{cjkk94,ame}. The analysis
of moments of the lepton momentum distribution, or the ratio of
charged vs. neutral lepton moments appear to be particularly
promising. 

The corrections are often sizable, in particular those to the decay rate. In
order to fix the scale in the running coupling constant and to gain
confidence in the numerical result, a calculation of NLO corrections
to the rate, if not the spectrum, is necessary.  Purely fermionic
loops have been considered in \cite{Smith,Cz2}.  
In the limit $m_t^2\gg m_W^2$ the result is particularly simple
\be
\Gamma_t &=& \Gamma_{\rm Born} \left[
1-{2\over 3} {\alpha_{\overline{\rm MS}}(m^2_t)\over \pi}
\left(4\zeta_2-{5\over 2}\right)
\right.
\nonumber\\
&&+
\left.
\left( {\alpha_{\overline{\rm MS}}\over \pi} \right)^2 
\left( -{2n_f\over 3} \right)
\left( {4\over 9} -{23\over 18}\zeta_2 - \zeta_3\right)
\right]
\ee
with 
\be
\Gamma_{\rm Born} = {G_F m_t^3 \over 8\sqrt{2}\pi  }
\ee
If we adopt the BLM prescription the large coefficient leads to a
large shift in the effective scale for $\alpha_s$: $\mu_{BLM} =
0.12m_t$. Similarly large correction factors have been observed
\cite{wise}
 for the
decay of $b$ into $l\nu$ plus a charmed or $u$ quark.

It should be emphasized that the magnitude of NLO corrections
$\sim(\alpha_s(m_b^2)/\pi)^2 \approx (0.07)^2$ is well comparable with
correction terms obtained in Heavy Quark Effective Theory ---
typically of order $(\Lambda/m_b)^2 \approx (0.05)^2$. Transitions at
zero recoil i.e. for the final state with $p_c = {m_c\over m_b} p_b$,
are particularly clean from the theoretical point of view. No
uncalculable form factor is present, allowing to determine $V_{cb}$
with remarkable precision. The first calculation of the full NLO QCD
corrections has therefore been performed at zero recoil \cite{czprl}.
Two
important simplifications are present in this case: 
\begin{itemize}
\item 
no real radiation
has to be considered, 
\item 
only relatively simple two loop integrals arise
which can be calculated in a series expansion. 
\end{itemize}

The resulting NLO corrections are smaller than the leading ones by
about a factor 4, reducing thus the theoretical error by a significant
factor.  Evidently these results can be considered a first important
step towards a complete NLO calculation of the heavy quark decay
rate.

\end{document}